\documentclass[runningheads]{llncs}
\usepackage{graphicx}
\usepackage{subfig}
\usepackage{xcolor,soul,framed} 
\usepackage{color}
\usepackage{enumitem}
\begin{document}
\title{TaNTIN: Terrestrial and Non-Terrestrial Integrated Networks-A collaborative technologies perspective for beyond 5G and 6G}

\author{Muhammad Waseem Akhtar\inst{1}\and
Syed Ali Hassan\inst{1} \and
}

\authorrunning{M.W. Akhtar et al.}

\institute{School of Electrical Engineering and Computer  Science (SEECS), National University of Sciences and Technology (NUST), Islamabad, Pakistan
\email{engr.waseemakhtar@seecs.edu.pk, ali.hassan@seecs.edu.pk}\\
$^*$ Corresponding authors: S. A.Hassan}
\maketitle              

\begin{abstract}
The world is moving toward globalization rapidly. Everybody has easy access to information with the spread of Internet technology. Businesses are growing beyond national borders. Internationalization affects every aspect of life. In this scenario, by dispersing functions and tasks across organizational borders, time and space, global organizations have higher requirements for collaboration. In order to allow decision-makers and knowledge workers, situated at different times and spaces, to work more efficiently, collaborative technologies are needed. In this paper, we give an overview of potential collaborative technologies, their benefits, risks and challenges, types, and elements. Based on the conceptualization of terrestrial and non-terrestrial integrated networks (TaNTIN), we highlight artificial intelligence (AI), blockchains, tactile Internet, mobile edge computing (MEC)/fog computing, augmented reality and virtual reality, and so forth as the key features to ensure quality-of-service (QoS) guarantee of futuristic collaborative services such as telemedicine, e-education, online gaming, online businesses, the entertainment industry. We also discuss how these technologies will impact human life in the near future. 

\keywords{\textcolor{black}{collaborative technologies\and blockchain\and 5G\and 6G\and e-learning\and telemedicine\and globalization\and video conferencing\and LMS\and CMS}}
\end{abstract}
\section{Introduction}\label{sec1}
Collaboration is a process through which individuals work together inside or outside of an organization to achieve certain goals effectively \cite{CNT02}. The parties concerned share information, objectives, and resources and make decisions together. Collaboration can be a powerful tool for achieving objectives and significantly increasing productivity. It is also a way of coordinating ideas for the generation of knowledge from different people. The demand for collaborative technologies has grown rapidly due to their potential to handle the dispersal of business activities, such as offshore business communications, need to work from home or remotely, online education, and remote interaction with customers. These technologies can improve business strategies by increasing profit, revenue, and market share by efficiently managing and streamlining complex business processes. The demand for collaborative network technologies has exponentially increased in the recent Coronavirus Disease 2019 (COVID-19) pandemic. The business leader, around the world, are emphasizing especially how the team of experts and decision-makers can work together more efficiently by leveraging collaborative technologies. \par
With the fast development of the Internet technologies, it has become easy to interconnect multiple applications. Collaborative technologies ensure that companies, corporations, and businesses can convene remote employees in virtual teams to execute a range of activities efficiently. Companies and staff from all around the world can interconnect, exchange their knowledge, and engage easily in project activities. The Internet or other means of technology are used to exchange information \cite{CNT03}. These collaboration strategies can reduce face-to-face interaction and improve efficiency if properly applied. The staff involved must gain teamwork expertise and be familiar with the processes. Collaboration familiarity is an individual’s willingness to express and integrate his or her thoughts with the team members to enhance the team’s performance.
The user plan provides a plethora of collaborative network applications. \textcolor{black}{A vast number of devices/internet-of-things (IoT) consisting of healthcare IoT, personal IoT, industrial IoT, and sensors on autonomous vehicles and plans are also included in the user plan. These devices and collaborative network applications generate a huge amount of data on a daily basis, ultimately triggering enhanced mobile broadband (eMBB), high spectral efficiency, massive machine-type communication (mMTC), high data rates and extremely reliable and low latency communication (URLLC) requirements \cite{akhtar}. The objective of the future communication network, i.e., beyond the fifth (5G)/sixth (6G) generation, is to meet all these requirements. Network inbox (NIB), Internet-of-everything (IoE), etc., are the terminologies used for future communications networks in the literature \cite{newone}.}\par
\textcolor{black}{In this paper, we introduce a novel terminology for a future communication network that is terrestrial and non-terrestrial integrated networks (TaNTIN). Conventional ground-based communication networks are identified by terrestrial networks, while non-terrestrial networks consist of unmanned aerial vehicles (UAV) and satellite communication systems, as well as maritime communication, space communication, and underground communication networks.} Blockchains, artificial intelligence (AI)/machine learning (ML), holography, haptics, augmented reality (AR)/virtual reality (VR), network slicing, network virtualization, smart IoTs, software-defined networks (SDN), cloudification/fog, etc., are the key enabling technologies that can be crucial to TaNTIN implementation. All these technologies are discussed in this paper and we focus on how these technologies will help TaNTIN to be realized with a wide range of collaborative network applications in B5G/6G systems.\par We start by discussing the architecture of TaNTIN, benefits, risks, and limitations of collaborative technologies in Section 2. Some key supporting and enabling technologies are discussed in Section 3. Section 4 gives an overview of elements of collaborating technologies and finally, we conclude our paper in Section 5.
\begin{figure*}[t]
\centerline{\includegraphics[scale=0.5]{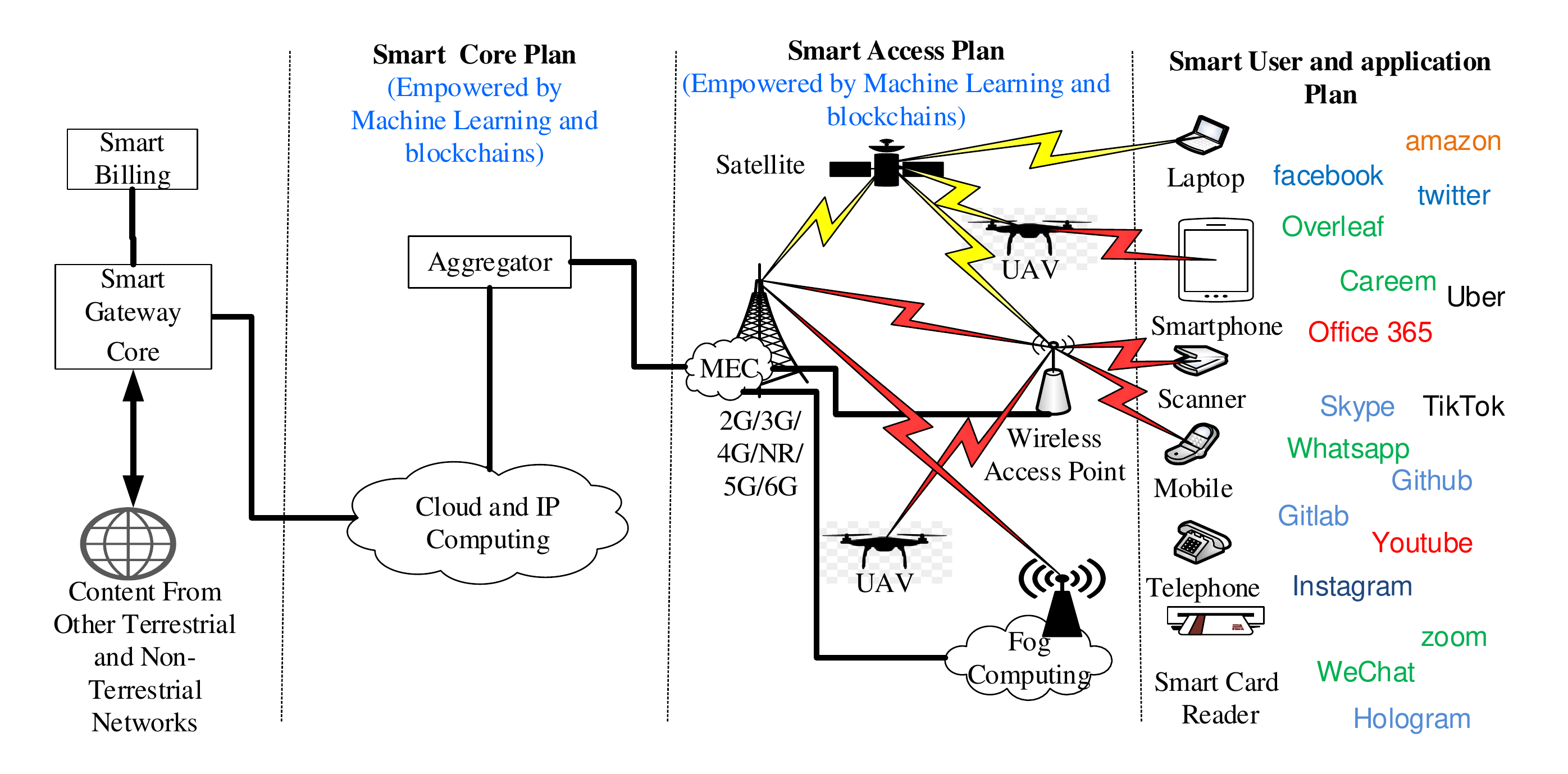}}
\caption{\textcolor{black}{An illustration of TaNTIN architecture for B5G/6G with collaborative network technologies perspective. The key enabling technologies such as blockchains, machine learning/artificial intelligence, MEC/fog computing, network virtualization/slicing/softwarization, etc., play the key roles in realization of collaborative technologies in future TaNTIN.} }
\label{6GArch}
\end{figure*}
\section{Architecture, Benefits, Risks, and Limitations}
In this section, we describe the architecture of TaNTIN, benefits, risks, and limitations. Fig.~\ref{6GArch} depicts the architecture of TaNTIN with a collaborative network technologies perspective. AI/ML-aided smart access plan ensures smooth communication through the network. \par
  Collaborating technologies have a lot of benefits for the participating entities. These advantages include improvement in companies' survival and the ability to achieve more common objectives, excellence in individual expertise, access to new/greater markets and expertise, exchange of new/greater markets capital, incorporating additional talents and strengths. \par
Our companies require dealing with individuals, clients, staff, and strategic partners. Employees and teams need to get the job done and pursue it. The question is how to ensure that the process goes smoothly and how to accomplish the job? That is why cooperation is crucial. With the enhancement of technologies, without any geographical obstacles, people from all over the world can communicate and share ideas through the Internet. Collaborative technologies allow valuable information to be shared by employees, partners, and clients. Also, the lack of resources and knowledge are the main limitations of collaborative network technologies.\par
It can present many challenges and risks if one relies only on collaborating network technologies.  The most obvious one is the time aspect. For example, in the case of an e-mail, it can take hours or days to wait for someone to read and reply to the message. In the meantime, valuable time and opportunities are lost. Miscommunication and misinterpretation can also result from e-mail. One sometimes needs to clear things out and nothing is accomplished by the time taken for the exchange of messages. It might reduce productivity. Therefore for our companies, we need a much better collaborative tool. Apart from this, security, privacy, and unemployment are the potential threats of collaborating network technologies. In the next section, we give an overview of potential network technologies for TaNTIN.
\section{Key supporting and enabling technologies}
In this section, we describe key enabling technologies for TaNTIN with the perspective of collaborative network technologies for the B5G/6G network.
\subsection{Augmented reality and virtual reality}
Besides, the simulated environment of virtual objects is aligned and superimposed on the physical world and is shared by users, so that during the collaborative meeting changes made to simulated objects are communicated to all members and observable immediately. The state-of-the-art networking technologies for remote connectivity, such as audio/video and application-sharing, help transcend the gap, showing the actual environments of remote parties.
\subsection{Network slicing}
Network operations generally work on the principle of simplicity and consistency. Because of huge numbers of connected devices, the network receives more and more data, and the networks are being overloaded. A telecommunication network can easily be compared to the vehicular traffic network in big cities both in terms of complexity and traffic volume. They both can also be managed similarly.\par 
The flow of data packets in the telecommunication network is similar to cars entering the cities both on streets and highways. However, the movement is not smooth. There is a situation when traffic is concentrated in sensitive places. The usual operation for maintaining the network cannot accommodate such a large amount of incoming data. Network slicing can help the smooth and unblocked flow of data traffic through the network in which each slice is a complete end-to-end network that provides a wide range of services if required.
\subsection{Network virtualization }
In computing, the idea of virtualization refers to machine resource abstraction. The most historic benefit of virtualization is the introduction of time-sharing systems that in turn make the use of physical resources more effective. In comparison, the separation principle is also used with virtualization implementation, which is viewed either as a fast deployment of package applications or as the runtime separation for parallel incompatible applications operating on the same physical infrastructure.\par
Virtualization made the operating system separate from the underlying hardware for the first time. This functionality opened the door to the portability of applications from hardware business to hardware. 
\subsection{Smart IoTs} 
IoT is known as a worldwide interconnected network of objects. Smart here indicates the capacity or ability of the IoT to respond to complex situations, for example, self-learning, and self-reliance. IoTs enhance the quality of living, to name just a few: at home, on the street, on the road, while ill, at work, while jogging, and at the fitness center. Healthcare IoTs can help monitor objects and persons (both employees and patients), recognize and authenticate users, and automatically capture the sensors' data. data collected provide real-time information to assist for medical diagnosis on patient health measurements. Complex computations and vast numbers of data transfers are needed for IoT devices which are not required within the standard sensor network \cite{CNTI01}. 
\subsection{Softwarization}
In the present day, software-defined networks (SDNs) are becoming a hot technology for the network's new paradigm. The network control is decoupled from data transmission. This change provides versatile scalability, programmability, high service capacity, and handy maintenance across the network system. Edge computing is being used to expand vehicle services by sharing computing activities across boundaries and local terminals. By integrating SDN with edge computing we can achieve the key performance indicators (KPI) of fifth-generation (5G)/ beyond 5G (B5G) networks.
\subsection{Cloudification}
We consider cloud computing to be a pay-per-use service that offers infrastructure, platforms, and software as a service  (IaaS, PaaS, and SaaS) \cite{CNTC01}. Cloud computing allows them to deter overprovision of information technology and training staff in small and medium-sized businesses. Therefore, small to medium-sized businesses will use a cloud if the Internet technology (IT) capability has to be improved. In general, for programs that are only available for a specific time, more resources are required. Becoming a cloud platform allowed businesses with massive IT networks, including Google and Amazon, to deliver their services respectively to the small to medium-sized industries, depending on pay-for-go and subscription models. 
\subsection{Blockchains}
While blockchain technology was proposed as the underlying Bitcoin technology in Nakamoto's Whitepaper in 2008, the principle was not frequently reported. Blockchain technology has a range of features that are evaluated concerning their interconnections and extract a selection of main features. For example, it is believed that both the attributes of "shared and public" and "low friction" improve system transparency because, without the control of a third party, the information would be made accessible freely to participants. When looking at the blockchain, two key aspects must be identified that evokes and decentralizes the trust. It encourages its decentralization by providing a private, secure, and scalable environment. \par
Blockchain technology allows participants to incorporate their own programs, create and distribute their own code to shape their own environment to create an open, diversified system. The so-called smart contract, a piece of code that holds as the programmed contractual agreement between two parties is an example of this feature.\par
In the next section, we highlight the elements of the collaborative technologies in TaNTIN.
\section{Elements of Collaborative Technologies}
In this section, we discuss the major elements in the collaborative network technologies that can play a part in the realization of these technologies. 
\subsection{\textcolor{black}{Mobile Phones and Laptops}}
The Internet has transformed our way of educating, studying, and gathering knowledge. The exponential spread of smartphone and social media technology has influenced the way whole generations think, interact, and function \cite{newtwo}. Increasingly, with the growth of the social technology revolution, our digital lifestyle is shifting. \par
Presently, higher education is forced to educate the masses of various students who demand both high-quality education and very interactive media.  As a consequence, it has become extremely difficult to teach this web-centric generation. The move of media giants is a more crucial indication that the Internet is heading to a web experience! , Google,  Apple Computer, Disney Internet Group, and Sony will gradually provide their smartphone content for online education to the students.
\subsection{Video Conferencing }
For most major businesses and organizations, the willingness of staff to work with remote partners is now important. The criteria for effective remote partnerships are established in decades of study. We observed that many aspects have contributed to remote collaboration \cite{CNTV02}. Quick and reliable visual real-time collaboration by video conferencing is combined with calendar applications, whereas laptops with video conferencing capacity are offered to most Google employees \cite{CNTV02}. \par
Video portals minimize connectivity behavioral costs and allow closely linked remote work to be effective for the first time. While there are extensive problems in distributed meetings by technology, the distance between collocated and remote interactions can be drastically minimized by a conjunction of collective culture with sufficient facilities, software, and data access.
\begin{table*}[t]
\caption{Collaborative tools/technologies used in different sectors of humans' life.}
\label{table1}
\centering 
\begin{tabular}{ |p{3.5cm}|p{2.7cm}|p{3.5cm}|p{4.5cm}| }
  \hline
\textbf{Education}& \textbf{Finance/businesses} &\textbf{Health} &\textbf{Entertainment/communication} \\
\hline
\hline
 \begin{itemize}[leftmargin=*] 
 \item {LMS, CMS}\item {GitHub, Git-lab} \item {Overleaf, Office 365} \item {Zoom, Skype, MS teams} \end{itemize} 
 &\begin{itemize}[leftmargin=*]
 \item {Amazone}\item {Alibaba, Draz } \item {Uber, Cream} \item {FoodPanda} \end{itemize} 
 & \begin{itemize}[leftmargin=*] 
 \item {Telemedicine} \item {Robotics}\item {Augmented/virtual reality} \item {Holography} \end{itemize}
 & \begin{itemize}[leftmargin=*] 
 \item {Facebook, Whatsapp, WeChat}\item {TikTok, Youtube, Instagram} \item {Waze, Tencent}\item {Zoom, Skype, MS teams} \end{itemize}
 \\
 \hline
\end{tabular} 
\end{table*}
\subsection{Online Gaming}
Video gaming is a means of developing our students' skills. Anthropologists, linguists, physicists, geographers, sociologists, psychologists, and others have the identifiable influence of new technology on social life \cite{CNTG01}. Collaborative online gaming environments can support several pedagogical techniques which can be used as both constructivist and instructor learning environments, for example, working together, students can work to their strengths, improve analytical reasoning ability and imagination, affirm their reasoning, and understand a wide array of different styles, talents, and interests. \par
Online gaming networks offer a similar forum for working together and learning from others. Studies of recreational users of large multi-user multiplayer role play games for example. The constructivist point of view also indicates that students learn further by exploring and experiencing their own interpretations from their encounters.
\subsection{Instant Chatting}
With the evolution of wireless technologies, users can access the Internet or hold video conferences on their cell phones or digital personal assistants (PDAs). These technologies offer tremendous potential, especially in education where not only the short message service (SMS), but also the Instant Messaging (IM) service is available. Mobile applications have now been spread, delivering billions of SMS every month, with 5.20 billion mobile users worldwide \cite{newth}. \par
The Internet's most popular software,  IM makes it so that users continue to remain linked to the site for a long time, and also aims to promote a more meaningful "online community" as no other technology has done before.  Instant Messenger, Skype, Google Talk, Yahoo Messenger, and Microsoft Network (MSN) Messenger are among the most popular IM applications. In addition to user icons, all IM systems support avatars (a moving symbol depicting a human in cyberspace or visuals of virtual reality). 
\subsection{E-education}
Various interactive software such as wikis, blogs, podcasts, overleaf, GitHub, Git-lab, learning management systems (LMS), campus management systems (CMS), e-mailing and virtual platforms are being employed. It also emerges as a new type of academic experience that brings social problems, history, ethics, and learner interaction more thoroughly into account in response to the needs of post-industrial society. This constant growth and the ongoing interplay of human, organizational and technical influences has complicated efforts to examine the contributions of different elements in e-learning and above all, our realization of the connection between technology and learning.\par
Table~\ref{table1} gives an overview of the tool and technologies used in various sectors of human life. These tools have dramatically shifted the way of working in a very positive and progressive direction. 
\section{Conclusion}
\textcolor{black}{We begin to move forward with the vision of cellular networking for beyond 5G/6G communications. With advantages such as improved employee efficiency in an organization, reduced operational costs and time, collaborative technologies have become an integral part of modern lifestyles. In this paper, we shed light on the architecture of next-generation wireless communication with a perspective to collaborative technologies. There is a strong correlation between the realization of collaborative technologies and 6G key performance indicators (KPI) such as URLLC, mMTC, and eMBB. We discussed the advantages, limitations, and risks of collaborative technology implementation. Towards the end, the key enabling technologies for terrestrial and non-terrestrial integrated networks (TaNTIN) and elements of collaborative technologies are elaborated in detail.}

\end{document}